\documentclass[12pt]{article}
\title{
The Parametrized Relativistic Particle and the Snyder space-time}
\author{Juan M. Romero\footnote{sanpedro@nucleares.unam.mx}
and J. David Vergara\footnote{vergara@nucleares.unam.mx} \\
\\
Instituto de Ciencias Nucleares, UNAM, \\
A. Postal 70-543, M\'exico D.F., M\'exico.}

\usepackage[dvips]{graphicx}

\usepackage[latin1]{inputenc}
\date{}
\begin{document}

\pagestyle{plain}

\maketitle

\begin{abstract}
Using the parametrized relativistic particle we obtain the
noncommutative Snyder space-time. In addition, we study the
consistency conditions between the boundary conditions and the
canonical gauges that give origin to noncommutative theories. Using
these results we construct a first order action, in the reduced
phase-space, for the Snyder particle with momenta fixed on the
boundary.
\\
\\
\\
\\
Pacs: 11.10.Nx, 02.40.Gh, 45.20.-d\\
\end{abstract}

\section{Introduction}

Recently it has been studied the possibility of consider
noncommutative space-times, i.e. spaces that satisfy relations of
the following type,
\begin{eqnarray}
[\hat X^{\mu},\hat X^{\nu}]=i\hbar \theta^{\mu\nu},
\end{eqnarray}
with $\theta^{\mu\nu}$ an antisymmetric tensor. For example, it is
possible to show in string theory that for some backgrounds, the low
energy limit of the theory implies field theories in noncommutative
spaces \cite{witten:gnus}. In this case $\theta^{\mu\nu}$ is a
constant tensor that implies violations of the Lorentz symmetry
\cite{carroll:gnus}.\\

However, long time ago H. Snyder \cite{sn:gnus} proposed a
noncommutative space-time, where $\theta^{\mu\nu}$  depends on the
space-time. This space is quite interesting since is discrete and
compatible with the Lorentz symmetry. Also, it is interesting to
mention that a space-time of the Snyder type it is obtained from the
quantum gravity in $2+1-$dimensions \cite{hooft:gnus}. This gives
indications that it is possible to regard that Quantum Gravity in
$3+1-$dimensions implies noncommutative spaces of the Snyder type.
Another interesting property of the Snyder space is that exist a
mapping between this space and the $\kappa-$Minkowski space-time
\cite{kowalsky:gnus}, and this space-time is one of the arenas of
the Doubly Special Relativity (DSR).

In the literature exist some realizations of the Snyder space, in
one of these realizations is used a theory with two times
\cite{nos:gnus}, another proposal \cite{girelli:gnus} is based
essentially on the ordinary free relativistic particle, see section
3. Other realization is based on the non-relativistic  particle with
interactions \cite{subir:gnus}. In this paper we propose a different
realization of the Snyder space-time, that is based in some way on a
combination of the proposals \cite{subir:gnus} and
\cite{girelli:gnus}. Our starting point is to use a generalization
of the Klein-Gordon equation proposed originally by Fock
\cite{fock:gnus}, rediscovered by  Stueckelberg, Nambu
\cite{St:gnus}, and Feynman \cite{fey:gnus}. In particular we use
the action of this particle in the massless case \cite{david:gnus}
and fixing a gauge we obtain a representation of the Snyder
space-time. The interesting points of this proposal are: It is a
more direct form to obtain the Snyder space-time, it is also
covariant from the beginning and allow us to construct an explicit
action for the particle in the momentum space. To understand more
clearly our proposal we start from the analysis of the
non-relativistic particle in Section 2, where we study the
consistency conditions that impose the boundary conditions to the
action. In Section 3 we introduce our realization and the action for
the particle in the momentum space-time.

\section{Parametrized non-relativistic particle}
The parametrized non-relativistic particle is the generally
covariant system obtained by including the non relativistic time $t$
among the dynamical variables. The action for paths obeying the
boundary conditions
\begin{equation}\label{nrpbc}
    x^{i}(\tau_1) = x^{i}_{1}, \ t(\tau_1)=t_{1}, \  x^{i}(\tau_2) = x^{i}_{2}, \ t(\tau_2)=t_{2},
\end{equation}
is given by
\begin{equation}\label{actnr}
    S_{nr}=\int_{\tau_1}^{\tau_2}d\tau\left(p_i\dot x^{i}+ p_t \dot t - \lambda
    \left( p_t +H({\bf p,x})\right)\right),
\end{equation}
where
\begin{equation}\label{cons1}
    \chi_1 =p_t +H({\bf p,x}),
\end{equation}
is a first class constraint, where $H({\bf p,x})$ is the Hamiltonian
of the system. To recover the dynamics of the relativistic particle
we usually impose a canonical gauge condition of the form
\begin{equation}\label{gc1nrp}
\chi_2=t-f(\tau)\approx 0,
\end{equation}
where $f(\tau)$ is fixed in such way that the boundary conditions
(\ref{nrpbc}) are satisfied, see \cite{he-dav}. Fixing the gauge
(\ref{gc1nrp}), the phase space action (\ref{actnr}) is reduced to
\begin{equation}\label{act1nr}
S_{nr}=\int_{\tau_1}^{\tau_2}d\tau\left(p_i\dot x^{i}-H({\bf
p,x})\dot f\right),
\end{equation}
and from the elimination of the parametrization we finally obtain
\begin{equation}\label{act2nr}
S_{nr}=\int_{f_1}^{f_2}df \left(p_i\frac{dx^{i}}{df}-H({\bf
p,x})\right).
\end{equation}
This is the action of the non-relativistic particle with time $f$
and boundary conditions
\begin{equation}\label{bc1nr}
    x^i(f_1)= x^i_1, \ \ \ x^i(f_2)=x^i_2.
\end{equation}
Now we want to construct noncommutative theories, by fixing gauge
conditions. Several authors (see for example \cite{stern:gnus}) have
been considered gauge conditions of the following form,
\begin{equation}\label{gc2}
   \chi_2= t+\theta^i p_i -f(\tau)\approx 0.
\end{equation}
The gauge condition (\ref{gc2}) is correct in the Dirac sense, since
the set of constraints $(\chi_1, \chi_2)$ forms a good set of second
class constraints and the Dirac brackets can be build. Now,
following Dirac's method for second class constraints
\cite{dirac:gnus}, we define the Dirac brackets. Given two phase
space functions $A$ and $B$ these brackets are given by
\begin{eqnarray}
\{A,B\}^{*}= \{A,B\}-\{A,\chi_{a}\}C^{ab}\{\chi_{b},B\},
\end{eqnarray}
with $C^{ab}$ the inverse matrix of  $C_{ab}=
\{\chi_{a},\chi_{b}\}$. The gauge condition (\ref{gc2})  implies the
following Dirac brackets for the canonical variables,
\begin{eqnarray}\label{nrdb1}
\{t,x^i\}^*&=&\frac{\theta^i}{1-\theta^k\frac{\partial H}{\partial
x^k}}, \ \ \ \ \{t,p_i\}^*=0, \ \ \ \
\{p_i,p_j\}^*=0, \\
\{x^i,x^j\}^*&=&\frac{\partial H}{\partial
p_i}\frac{\theta^j}{1-\frac{\partial H}{\partial
x^k}\theta^k}-\frac{\theta^i}{1-\frac{\partial H}{\partial
x^k}\theta^k}\frac{\partial H}{\partial p_j},\ \
\{x^i,p_j\}^*=\delta^i_j+\frac{\theta^i\frac{\partial H}{\partial
x^j}}{1-\theta^k\frac{\partial H}{\partial x^k}}. \nonumber
\end{eqnarray}
However, to impose this gauge condition in the action (\ref{actnr}),
it does not have sense. The reason of that is the fact that we will
obtain an action with no well defined boundary conditions. For
example, we will get the simultaneous fixing of the coordinates
$x^i$ and the momenta $p_i$. Then, we will need an action where from
the beginning we fix variables on the boundary that are consistent
with the gauge condition. In this case, we see that a possibility is
to fix on the boundary the variables $(t,p_i)$ since this set is a
complete set of commuting variables, according to the Dirac brackets
(\ref{nrdb1}), then the respective action is given by,
\begin{equation}\label{actnr4}
    S_{nr1}=\int_{\tau_1}^{\tau_2}d\tau\left(-x^{i}\dot p_i + p_t \dot t - \lambda
    \left( p_t +H(\bf{p},\bf{x})\right)\right).
\end{equation}
Now fixing the gauge (\ref{gc2}) and with the elimination of the
parametrization we get finally,
\begin{equation}\label{actnr5}
 S_{nr1}=\int_{f_1}^{f_2}df\left((\theta^i
 H({\bf p,x})-x^{i}) \frac{dp_i}{df}- H({\bf{p},\bf{x}})\right),
\end{equation}
with boundary conditions,
\begin{equation}\label{bc2nr}
    p_i(f_1)= p_{i1}, \ \ \ p_i(f_2)=p_{i2}.
\end{equation}
The action (\ref{actnr5}) with the boundary conditions (\ref{bc2nr})
is a well defined physical problem that we can quantize using for
example the path integral. So we learnt that to fix gauge conditions
of  the type (\ref{gc2}), we must be careful that the gauge
condition be consistent with the boundary conditions, and in
consequence with the base selected to quantize the theory.

\section{Snyder space-time}
To obtain a realization of the Snyder space-time in $d+1$-dimensions
is useful to start from a space-time with one extra dimension. For
example in the Ref. \cite{girelli:gnus} it was shown that the Snyder
space-time appears by imposing a particular gauge condition to the
action
\begin{eqnarray}
S=\int d\tau \left( \dot
X_{M}P^{M}-\frac{\lambda_{1}}{2}(P^{M}P_{M}-\kappa^{2})
-\lambda_{2}(P_{4}- M)\right), \label{eq:gi}
\end{eqnarray}
where $X^{M}=(X^{\mu},X^{4})$, and  $P^{M}=(P^{\mu},P^{4}),$ with
$\mu=0,...3$. Furthermore, the background metric is flat with
signature $sig(\eta)=(-,+,...,+).$ It is interesting to observe that
the previous action is directly equivalent to the relativistic
particle, we can see this from the elimination in the the action
(\ref{eq:gi}) of the auxiliary momenta  $P_4$ and the Lagrange's
multiplier $\lambda_2$ associated to the second constraint, we
obtain
\begin{equation}\label{act2}
S=\int \left(\dot X^\mu P_\mu
-\frac{\lambda_{1}}{2}(P^{\mu}P_{\mu}-\tilde m^{2})+M\dot
X_{4}\right), \qquad \tilde m=\sqrt{\kappa^{2}-M^2}
\end{equation}
%
%\begin{eqnarray}
%S=\int d\tau \left(-\tilde m \sqrt{-\dot X_{\mu}\dot X^{\mu}} +M\dot
%X_{4}\right), . \label{eq:parl}
%\end{eqnarray}
%
Thus, we obtain the action of a free relativistic particle with mass
$\tilde m$, plus a total derivative of the extra dimension that does
not interact with the additional variables. We also see from
(\ref{act2}), that from the definition of the momentum  for the
$X_4$ variable we get immediately the second constraint of
(\ref{eq:gi}) and we recover this action.

Now our propose is to obtain another realization of the Snyder
space-time where the extra dimension plays different role in such
way we can extract some physical content of this new dimension. To
that end we start from the action
\begin{eqnarray}\label{act3}
S=\int d\tau \left( \frac{\dot X_{\mu}\dot X^{\mu}}{2\dot \eta}
 \right).
\end{eqnarray}
Here the extra dimension is given by the parameter $\eta$. We will
assume that this parameter is an invariant scalar under Lorentz
transformations of the $3+1$-dimensional space-time
\cite{david:gnus}.

The equations of motion of this system are
\begin{eqnarray}
\frac{d}{d\tau}\left(\frac{\dot X_{\mu}}{\dot \eta }\right)&=&0,\\
\frac{d}{d\tau}\left(\frac{\dot X_{\mu} \dot X^{\mu}}{2\dot \eta^{2}}\right )&=&0.
\end{eqnarray}
From these equations we get,
\begin{eqnarray}
\frac{\dot X_{\mu}}{\dot \eta }&=&c_{\mu},\qquad c_{\mu}={\rm constant},\label{condi0}\\
\frac{\dot X_{\mu} \dot X^{\mu}}{2\dot \eta^{2}}
&=&\frac{c_{\mu}c^{\mu}}{2}.\label{condi1}
\end{eqnarray}
In this case the canonical momenta are given by
\begin{eqnarray}
P_{\eta}&=&-\frac{\dot X_{\mu} \dot X^{\mu}}{2\dot \eta^{2}},\\
P_{\mu}&=&\frac{\dot X_{\mu}}{\dot \eta }.\label{defmome}
\end{eqnarray}
From the definition of these momenta we obtain the first class
constraint
\begin{eqnarray}\label{con3}
\phi= P_{\eta}+\frac{1}{2} P_{\mu} P^{\mu} =0.
\end{eqnarray}
This means that the physical states of the theory in the base
$|x^\mu, \eta\rangle$ satisfy the following quantum evolution
equation
\begin{eqnarray}
\left (\hat P_{\eta}+\frac{1}{2} \hat P_{\mu} \hat P^{\mu}
\right)\psi =0, \quad {\rm with}\quad \hat P_{\eta}=-i\hbar
\partial_{\eta}, \quad  \hat P_{\mu}=-i\hbar \partial_{\mu}.
\label{eq:kgsm}
\end{eqnarray}
This expression is a generalization of the Klein-Gordon equation,
that was originally proposed by V. Fock \cite{fock:gnus}, and also
used by Stueckelberg and Nambu \cite{St:gnus}. An interesting
property of this system is that integrating his propagator over the
$\eta$ parameter we recover the usual propagator of the Klein-Gordon
equation \cite{fey:gnus}. Furthermore, the extra dimension $\eta$
gives an alternative mechanism to give mass to the particle
\cite{Fanchi:gnus}. Solving this equation for $\eta$ we can obtain
the Klein-Gordon equation, but with the characteristic that the mass
$m^2$ could be positive or negative so we also have tachyons. Is
interesting to notice that for a given time dependent potential
$V(X^0)$ we can obtain solutions to the equation (\ref{eq:kgsm})
where the mass is time dependent, and these models have been used as
toy models to
analyze spacelike singularities in string theory \cite{Silver}.\\

Another form to see these facts is to consider a gauge condition to
the system. The natural one is of the form,
\begin{eqnarray}\label{con2}
\chi=\eta-\tau\approx0.
\end{eqnarray}
In this case the equations of motion are the following
\begin{eqnarray}
\ddot X_{\mu}&=&0,\\
\dot X_{\mu} \dot X^{\mu}&=&c^{2}. \label{eq:rest}
\end{eqnarray}
For $c^{2}=0$, we have the equations of motion of a massless
relativistic. For $c^{2}<0$, we get a relativistic particle and for
$c^{2}>0$, we obtain a tachyon.\\

The canonical form of the action (\ref{act3}), consistent with the
gauge condition (\ref{con2}) is given by
\begin{equation}\label{actca}
S=\int d\tau \left( P_\eta \dot \eta +P_\mu \dot X^\mu -\lambda
\left(P_{\eta}+\frac{1}{2}  P_{\mu}
P^{\mu}\right)\right).
\end{equation}
To obtain a noncommutative theory taking as starting point the
action (\ref{actca}) we follow the procedure established in
\cite{stern:gnus}, where using a selected gauge condition is
possible to obtain a noncommutative theory from the Dirac brackets.
However, we need to take into account that the boundary conditions
must be consistent with the new gauge condition. This fact can imply
a modification of the kinetic term of the action, as we saw in the
case of the non-relativistic particle.

To get the Snyder space-time we propose the following gauge
condition
\begin{eqnarray}
\chi_{2}=f(\tau)+\eta-\alpha X^{\mu}P_{\mu}\approx 0, \qquad
\alpha={\rm constant}. \label{eq:condicion}
\end{eqnarray}
We must notice that this gauge condition is not unique, but the
other possible choices are related by a canonical transformation.
Furthermore, the gauge condition (\ref{eq:condicion}) is a good
canonical gauge condition in the Dirac sense since, if we define
$\chi_{1}=\phi,$ we obtain
\begin{eqnarray}
\{\chi_{1},\chi_{2}\}=-\left[1+2\alpha P_{\eta}\right].
\end{eqnarray}
Then, the constraints $(\chi_1, \chi_2)$ form a good set of
canonical second class constraints.  In particular, for the Dirac
brackets of the reduced phase space variables  $X_{\mu}$ and
$P_{\nu}$ we obtain
\begin{eqnarray}
\{X_{\mu},X_{\nu}\}^{*}&=&\frac{1}{l^{2}}L_{\mu\nu}, \qquad L_{\mu\nu}=
X_{\mu}P_{\nu}-P_{\mu}X_{\nu},\label{cxx}\\
\{X_{\mu},P_{\nu}\}^{*}&=&\eta_{\mu\nu}+\frac{1}{l^{2}}P_{\mu}P_{\nu},\label{cxp}\\
\{P_{\mu},P_{\nu}\}^{*}&=&0\label{cpp},
\end{eqnarray}
where
\begin{equation}\label{L2}
 l^{2}=\frac{1}{\alpha}\left( 1+2 \alpha P_{\eta}\right)=
\frac{1}{\alpha}\left( 1-\alpha P_{\mu}P^{\mu}\right).
\end{equation}
Using these brackets we can compute the Dirac brackets for $\eta$
and $P_\eta$ using the constraint (\ref{con3}) and the gauge
condition (\ref{eq:condicion}). In particular we obtain that
$P_\eta$ commute with all the momenta $P_\mu$. Now, to quantize this
theory we must promote the Dirac brackets to commutators and we get
for the coordinates of the phase space a realization of the Snyder
space-time \cite{sn:gnus}. We must notice that in the usual
formulation of Snyder we have that $l$ is a constant and is in some
sense a minimal distance to which the space-time is discrete. In our
case, $l$ depends of the additional momentum $P_\eta$ or using  in
strong way the constraints, in the momenta of the particle, see
(\ref{L2}), and it is not in the center of the algebra. However, $l$
commutes with,
\begin{eqnarray}
\{l^{2},L_{\mu\nu}\}^{*}=\{l^{2},P_{\mu}P_{\nu}\}^{*}=0.
\end{eqnarray}
Then, we do not have ordering problems in the commutators
(\ref{cxx})-(\ref{cpp}). Now to construct the action of this
particle we need to fix the appropriated boundary conditions that
are consistent with the Dirac brackets (\ref{cxx})-(\ref{cpp}) and
in consequence with the constraint (\ref{con3}) and the gauge
condition (\ref{eq:condicion}). From this brackets we see that the
momenta $(P_\eta,P_\mu)$ form a complete set of commuting variables
of the extended phase space, then we can fix these variables on the
boundary. The corresponding action will be in this case,
\begin{equation}\label{act7}
S_{sp}=\int^{\tau_2}_{\tau_1} d\tau \left(-\eta \dot P_\eta-X^\mu
\dot P_\mu -\lambda\left(P_{\eta}+\frac{1}{2}  P_{\mu}
P^{\mu}\right) \right).
\end{equation}

Now, imposing in strong way the second class constraints we obtain
the action in the reduced phase space
\begin{equation}\label{act8}
S_{rsp}=\int^{\tau_2}_{\tau_1} d\tau \left( X^\mu\left(\alpha P_\mu
P^\beta -\delta_\mu^\beta \right)\dot P_{\beta}-f(\tau)P_\mu \dot
P^\mu\right).
\end{equation}
with boundary conditions,
\begin{equation}\label{bc8}
P_\mu(\tau_1)=P_{\mu 1}, \ \ \ P_\mu(\tau_2)=P_{\mu 2}.
\end{equation}
This is a first order action for the free relativistic particle in
the Snyder space-time. From the variation of $X^\mu$ we get that the
particle  must satisfy,
\begin{equation}\label{eqm}
\left(\alpha P_\mu P^\beta -\delta_\mu^\beta \right)\dot
P_{\beta}=0.
\end{equation}
Using the fact that $\left(\alpha P_\mu P^\beta -\delta_\mu^\beta
\right)$ is an invertible matrix we get as result the equation of
motion of the free relativistic particle $\dot P_\beta=0$. It is
easy to see that this equation is consistent with the equation
(\ref{condi0}) and the definition of the momentum (\ref{defmome}).

In conclusion, in this paper we analyze the consistency condition
between the boundary condition and the gauge conditions used to
obtain noncommutative theories. This analysis impose strong
conditions in the kinetic term of the Hamiltonian action and can
avoid the elimination of all the momenta to construct a second order
action in the reduced space. Furthermore this procedure is
consistent with the election of a complete base of commuting
observables at the level of quantum mechanics. As a second
interesting point of our article we construct using the massless
parametrized relativistic particle an action for a particle in the
Snyder space-time. We check the consistency conditions between the
boundary conditions and the gauge condition. In this way we obtain a
first order action for the Snyder particle in the reduced
phase-space and in consequence this action is consistent with the
second order action of the parametrized relativistic particle. So
the main result of our paper is that the more useful way to analyze
the Snyder particle is to use the action of the massless
parametrized relativistic particle.

\section*{Acknowledgments}
One of the authors acknowledge partial support from CONACyT project
47211-F(J.D.V.) and  DGAPA-UNAM grant IN104503 (J.D.V.).

\end{document}